\documentclass[12pt,journal,onecolumn]{IEEEtran}

\usepackage{amssymb,amsmath,amsfonts,bm,epsfig,graphicx,theorem,latexsym}
\usepackage{rotating,setspace,latexsym,epsf,color,subfigure}
\usepackage{cite}
\newcommand{\Ev}{\mathbf{E}}
\newcommand{\tv}{\mathbf{t}}

\newcommand{\rv}{\mathbf{r}}
\newcommand{\lv}{\mathbf{l}}

\newcommand{\lambdav}{\boldsymbol{\lambda}}
\newcommand{\gammav}{\boldsymbol{\gamma}}

\newtheorem{Theorem}{Theorem}
\newtheorem{Corollary}{Corollary}
\newtheorem{Lemma}{Lemma}

\newtheorem{Definition}{Definition}
\newenvironment{Proof}[1]{\medskip\par\noindent
{\bf Proof:\,}\,#1}{{\mbox{\,$\blacksquare$}\par}}

\makeatletter

\begin{document}
\IEEEoverridecommandlockouts

\title{Broadcasting with an Energy Harvesting Rechargeable Transmitter\thanks{This
work was supported by NSF Grants CCF 04-47613, CCF 05-14846,
CNS 07-16311, CCF 07-29127, CNS 09-64632.}}

\author{Jing Yang \qquad Omur Ozel \qquad Sennur Ulukus \\
\normalsize Department of Electrical and Computer Engineering \\
\normalsize University of Maryland, College Park, MD 20742 \\
\normalsize {\it yangjing@umd.edu} \qquad {\it omur@umd.edu} \qquad {\it ulukus@umd.edu} }

\maketitle
\thispagestyle{plain}
\pagestyle{plain}
\setcounter{page}{1}
\pagenumbering{arabic}

\begin{abstract}
In this paper, we investigate the {\it transmission completion
time} minimization problem in a two-user additive white
Gaussian noise (AWGN) broadcast channel, where the transmitter
is able to harvest energy from the nature, using a rechargeable
battery. The harvested energy is modeled to arrive at the
transmitter randomly during the course of transmissions. The
transmitter has a fixed number of packets to be delivered to
each receiver. Our goal is to minimize the time by which all of
the packets for both users are delivered to their respective
destinations. To this end, we optimize the transmit powers and
transmission rates intended for both users. We first analyze
the structural properties of the optimal transmission policy.
We prove that the optimal {\it total} transmit power has the
same structure as the optimal single-user transmit power
\cite{ciss10, tcom-submit}. We also prove that there exists a
{\it cut-off} power level for the stronger user. If the optimal
total transmit power is lower than this cut-off level, all
transmit power is allocated to the stronger user, and when the
optimal total transmit power is larger than this cut-off level,
all transmit power above this level is allocated to the weaker
user. Based on these structural properties of the optimal
policy, we propose an algorithm that yields the globally
optimal off-line scheduling policy. Our algorithm is based on
the idea of reducing the two-user broadcast channel problem
into a single-user problem as much as possible.
\end{abstract}

\newpage

\section{Introduction}
We consider a wireless communication network where users are
able to harvest energy from the nature using rechargeable
batteries. Such energy harvesting capabilities will make
sustainable and environmentally friendly deployment of wireless
communication networks possible. While energy-efficient
scheduling policies have been well-investigated in traditional
battery powered (un-rechargeable) systems \cite{acm_2002,
modiano_calculus, modiano_fading, ws_chen07, infocom_2002,
it_2004}, energy-efficient scheduling in energy harvesting
networks with nodes that have rechargeable batteries has only
recently been considered \cite{ciss10, tcom-submit}. References
\cite{ciss10, tcom-submit} consider a single-user communication
system with an energy harvesting transmitter, and develop a
packet scheduling scheme that minimizes the time by which all
of the packets are delivered to the receiver.

In this paper, we consider a multi-user extension of the work
in \cite{ciss10, tcom-submit}. In particular, we consider a
wireless broadcast channel with an energy harvesting
transmitter. As shown in Fig.~\ref{fig:bc}, we consider a
broadcast channel with one transmitter and two receivers, where
the transmitter node has three queues. The data queues store
the data arrivals intended for the individual receivers, while
the energy queue stores the energy harvested from the
environment. Our objective is to adaptively change the
transmission rates that go to both users according to the
instantaneous data and energy queue sizes, such that the total
{\it transmission completion time} is minimized.

\begin{figure}[p]
\begin{center}
\scalebox{0.65} {\epsffile{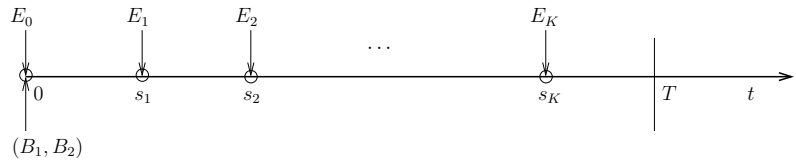}}
\end{center}
\vspace*{-0.1in}
\caption{System model. $(B_1,B_2)$ bits to be transmitted to users are available at the transmitter
at the beginning. Energies arrive (are harvested) at points denoted by $\circ$. $T$ denotes the transmission completion time by which all of the bits are delivered to their respective destinations.}
\label{fig:bc_system}
\vspace*{0.55in}
\end{figure}

In this paper, we focus on finding the optimum {\it off-line}
schedule, by assuming that the energy arrival profile at the
transmitter is known ahead of time in an off-line manner, i.e.,
the energy harvesting times and the corresponding harvested
energy amounts are known at time $t=0$. We assume that there
are a total of $B_1$ bits that need to be delivered to receiver
1, and $B_2$ bits that need to be delivered to receiver 2,
available at the transmitter at time $t=0$. As shown in
Fig.~\ref{fig:bc_system}, energy arrives (is harvested) at
points in time marked with $\circ$; in particular, $E_{k}$
denotes the amount of energy harvested at time $s_k$. Our goal
is to develop a method of transmission to minimize the time,
$T$, by which all of the data packets are delivered to their
respective receivers.

The optimal packet scheduling problem in a single-user energy
harvesting communication system is investigated in
\cite{ciss10, tcom-submit}. In \cite{ciss10, tcom-submit}, we
prove that the optimal scheduling policy has a ``majorization''
structure, in that, the transmit power is kept constant between
energy harvests, the sequence of transmit powers increases
monotonically, and only changes at some of the energy
harvesting instances; when the transmit power changes, the
energy constraint is tight, i.e., at the times when the
transmit power changes, the total consumed energy equals the
total harvested energy. In \cite{ciss10, tcom-submit}, we
develop an algorithm to obtain the optimal off-line scheduling
policy based on these properties. Reference \cite{kaya_yener}
extends \cite{ciss10, tcom-submit} to the case where
rechargeable batteries have finite sizes. We extend
\cite{ciss10, tcom-submit} in \cite{energy_fading} to a fading
channel.

References \cite{kaya_yener, energy_fading} investigate two
related problems. The first problem is to maximize the
throughput (number of bits transmitted) with a given deadline
constraint, and the second problem is to minimize the
transmission completion time with a given number of bits to
transmit. These two problems are ``dual'' to each other in the
sense that, with a given energy arrival profile, if the maximum
number of bits that can be sent by a deadline is $B^*$ in the
first problem, then the minimum time to transmit $B^*$ bits in
the second problem must be the deadline in the first problem,
and the optimal transmission policies for these two problems
must be identical. In this paper, we will follow this ``dual
problems'' approach. We will first attack and solve the first
problem to determine the structural properties of the optimal
solution. We will then utilize these structural properties to
develop an iterative algorithm for the second problem. Our
iterative approach has the goal of reducing the two-user
broadcast problem into a single-user problem as much as
possible, and utilizing the single-user solution in
\cite{ciss10, tcom-submit}. The second problem is also
considered in the independent work \cite{uysal_paper} which
uses convex optimization techniques to reduce the problem into
local sub-problems that consider only two energy arrival
epochs at a time.

We first analyze the structural properties of the optimal
policy for the first problem where our goal is to maximize the
number of bits delivered to both users under a given deadline
constraint. To that end, we first determine the {\it maximum
departure region} with a given deadline constraint $T$. The
maximum departure region is defined as the set of all $(B_1,
B_2)$ that can be transmitted to users reliably with a given
deadline $T$. In order to do that, we consider the problem of
maximizing $\mu_1B_1+\mu_2 B_2$ under the energy causality
constraints for the transmitter, for all $\mu_1, \mu_2 \geq 0$.
Varying $\mu_1$, $\mu_2$ traces the boundary of the maximum
departure region. We prove that the optimal {\it total}
transmit power policy is independent of the values of $\mu_1$,
$\mu_2$, and it has the same ``majorization'' structure as the
single-user non-fading solution. As for the way of splitting
the total transmit power between the two users, we prove that
there exists a {\it cut-off} power level for the stronger user,
i.e., only the power above this {\it cut-off} power level is
allocated to the weaker user.

We then consider the second problem, where our goal is to
minimize the time, $T$, by which a given $(B_1,B_2)$ number of
bits are delivered to their intended receivers. As discussed,
since the second problem is ``dual'' to the first problem, the
optimal transmission policy in this problem has the same
structural properties as in the first problem. Therefore, in
the second problem as well, there exists a {\it cut-off} power
level. The problem then becomes that of finding an optimal {\it
cut-off} power such that the transmission times for both users
become identical and minimized. With these optimal structural
properties, we develop an iterative algorithm that finds the
optimal schedule efficiently. In particular, we first use the
fact that the optimum total transmit power has the same
structural properties as the single-user problem, to obtain the
first optimal total power, $P_1$, i.e., the optimal total power
in the first epoch. Then, given the fact that there exists a
{\it cut-off} power level, $P_c$, for the stronger user, the
optimal transmit strategy depends on whether $P_1$ is smaller
or larger than $P_c$, which, at this point, is unknown.
Therefore, we have two cases to consider. If $P_c$ is smaller
than $P_1$, then the stronger user will always have a constant,
$P_c$, portion of the total transmit power. This reduces the
problem to a single-user problem for the second user, together
with a fixed-point equation in a single variable ($P_c$) to be
solved to ensure that the transmissions to both users end at
the same time. On the other hand, if $P_c$ is larger than
$P_1$, this means that all of $P_1$ must be spent to transmit
to the first (stronger) user. In this case, the number of bits
delivered to the first user in this time duration can be
subtracted from the total number of bits to be delivered to the
first user, and the problem can be started anew with the
updated number of bits $(B_1,B_2')$ after the first epoch.
Therefore, in both cases, the broadcast channel problem is
essentially reduced to single-user problems, and the approach
in \cite{ciss10, tcom-submit} is utilized recursively to solve
the overall problem.

\section{System Model and Problem Formulation}
\begin{figure}[p]
\begin{center}
\scalebox{0.6} {\epsffile{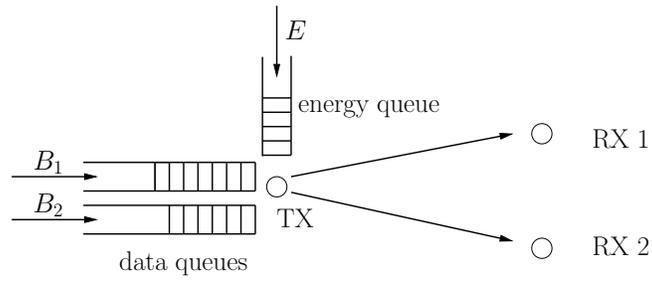}}
\end{center}
\vspace*{-0.1in}
\caption{An energy harvesting two-user broadcast channel.}
\label{fig:bc}
\vspace*{0.55in}
\end{figure}

\begin{figure}[p]
\begin{center}
\scalebox{0.5} {\epsffile{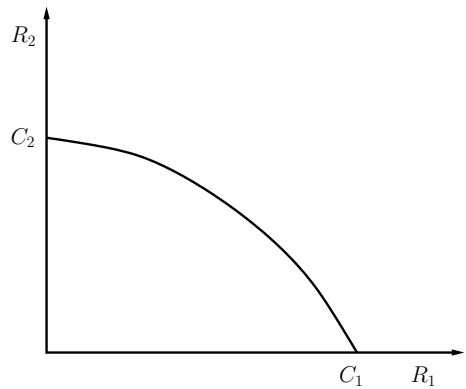}}
\end{center}
\vspace*{-0.1in}
\caption{The capacity region of the two-user AWGN broadcast channel.}
\label{fig:bc_capacity}
\vspace*{0in}
\end{figure}

The system model is as shown in Figs.~\ref{fig:bc}
and~\ref{fig:bc_system}. The transmitter has an energy queue
and two data queues (Fig.~\ref{fig:bc}). The physical layer is
modeled as an AWGN broadcast channel, where the received
signals at the first and second receivers are
\begin{align}
Y_1&=X+Z_1\\
Y_2&=X+Z_2
\end{align}
where $X$ is the transmit signal, and $Z_1$ is a Gaussian noise
with zero-mean and unit-variance, and $Z_2$  is a Gaussian
noise with zero-mean and variance $\sigma^2$, where
$\sigma^2>1$. Therefore, the second user is the {\it degraded} (weaker)
user in our broadcast channel. Assuming that the transmitter
transmits with power $P$, the capacity region for this two-user
AWGN broadcast channel is \cite{cover}
\begin{align}
  r_1&\leq \frac{1}{2}\log_2\left(1+\alpha P\right)\\
  r_2&\leq \frac{1}{2}\log_2\left(1+\frac{(1-\alpha) P}{\alpha P+\sigma^2}\right)\label{eqn:rate2}
\end{align}
where $\alpha$ is the fraction of the total power spent for the
message transmitted to the first user. Let us denote
$f(p)\triangleq \frac{1}{2}\log_2\left(1+p\right)$ for future
use. Then, the capacity region is $r_1\leq f(\alpha P)$,
$r_2\leq f\left(\frac{(1-\alpha) P}{\alpha P+\sigma^2}\right)$.
This capacity region is shown in Fig.~\ref{fig:bc_capacity}.

Working on the boundary of the capacity region, we have
\begin{align}
  P&=2^{2(r_1+r_2)}+(\sigma^2-1)2^{2r_2}-\sigma^2\\
  &\triangleq g(r_1,r_2)\label{dfn:g}
\end{align}
As shown in Fig.~\ref{fig:bc}, the transmitter has $B_1$ bits
to transmit to the first user, and $B_2$ bits to transmit to
the second user. Energy is harvested at times $s_k$ with
amounts $E_{k}$. Our goal is to select a transmission policy
that minimizes the time, $T$, by which all of the bits are
delivered to their intended receivers. The transmitter adapts
its transmit power and the portions of the total transmit power
used to transmit signals to the two users according to the
available energy level and the remaining number of bits. The
energy consumed must satisfy the causality constraints, i.e.,
at any given time $t$, the total amount of energy consumed up
to time $t$ must be less than or equal to the total amount of
energy harvested up to time $t$.

Before we proceed to give a formal definition of the
optimization problem and propose the solution, we start with
the ``dual'' problem of this transmission completion time
minimization problem, i.e., instead of trying to find the
minimal $T$, we aim to identify the maximum number of bits the
transmitter can deliver to both users by any fixed time $T$. As
we will observe in the next section, solving the ``dual''
problem enables us to identify the optimal structural
properties for both problems, and these properties eventually
help us reduce the original problem into simple scenarios,
which can be solved efficiently.

\section{Characterizing $\mathcal{D}(T)$: Largest $(B_1, B_2)$ Region for a Given $T$}
In this section, our goal is to characterize the maximum
departure region for a given deadline $T$. We define it as
follows.
\begin{Definition}
For any fixed transmission duration $T$, the maximum departure
region, denoted as $\mathcal{D}(T)$, is the union of $(B_1,
B_2)$ under any feasible rate allocation policy over the
duration $[0,T)$, i.e.,
$\mathcal{D}(T)=\bigcup_{r_1(t),r_2(t)}(B_1,B_2)(r_1(t),r_2(t))$,
subject to the energy constraint $\int_{0}^t g(r_1,r_2)(\tau)
d\tau\leq \sum_{i:s_i<t}E_i$, for $0\leq t \leq T$.
\end{Definition}

We call any policy which achieves the boundary of
$\mathcal{D}(T)$ to be optimal. In the single-user scenario in
\cite{ciss10}, we first examined the structural properties of
the optimal policy. Based on these properties, we developed an
algorithm to find the optimal scheduling policy. In this
broadcast scenario also, we will first analyze the structural
properties of the optimal policy, and then obtain the optimal
solution based on these structural properties. The following
lemma which was proved for a single-user problem in
\cite{ciss10, tcom-submit} was also proved for the broadcast
problem in \cite{uysal_paper}.

\begin{Lemma}\label{lemma:const}
Under the optimal policy, the transmission rate remains constant between energy harvests,
i.e., the rate only potentially changes at an energy harvesting
epoch.
\end{Lemma}
\begin{Proof}
We prove this using the strict convexity of $g(r_1,r_2)$. If
the transmission rate for any user changes between two energy
harvesting epochs, then, we can always equalize the
transmission rate over that duration without contradicting with
the energy constraints. Based on the convexity of $g(r_1,r_2)$,
after equalization of rates, the energy consumed over that
duration decreases, and the saved energy can be allocated to
both users to increase the departures. Therefore,
changing rates between energy harvests is sub-optimal.
\end{Proof}

\begin{figure}[p]
\begin{center}
\scalebox{0.4} {\epsffile{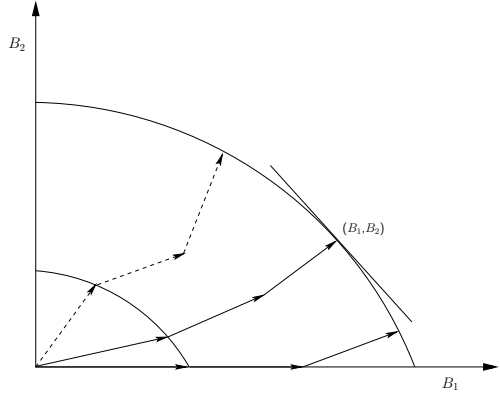}}
\end{center}
\vspace*{-0.1in}
\caption{The maximum departure region and trajectories to reach the boundary.
Dotted trajectory is not possible.}
\label{fig:trajectory}
\vspace*{-0.1in}
\end{figure}

Therefore, in the following, we only consider policies where
the rates are constant between any two consecutive energy
arrivals. We denote the rates that go to both users as
$(r_{1n},r_{2n})$ over the duration $[s_{n-1},s_n)$.
With this property, an illustration of the maximum departure region is shown in Fig.~\ref{fig:trajectory}.

\begin{Lemma}
$\mathcal{D}(T)$ is a convex region.
\end{Lemma}
\begin{Proof}
Proving the convexity of $\mathcal{D}(T)$ is equivalent to
proving that, given any two achievable points $(B_1,B_2)$ and
$(B_1', B_2')$ in $\mathcal{D}(T)$, any point on the line between these two points
is also achievable, i.e., in $\mathcal{D}(T)$. Assume that $(B_1,B_2)$ and $(B_1', B_2')$
can be achieved with rate allocation policies $(\rv_1,\rv_2)$
and $(\rv_1',\rv_2')$, respectively. Consider the policy
$(\lambda \rv_1+\bar{\lambda}\rv_1',\lambda
\rv_2+\bar{\lambda}\rv_2' )$, where $\bar{\lambda}=1-\lambda$.
Then, the energy consumed up to $s_n$ is
\begin{align}
    \sum_{i=1}^ng(\lambda r_{1i}+\bar{\lambda}r_{1i}',\lambda r_{2i}+
    \bar{\lambda}r_{2i}')l_i &\leq \lambda \sum_{i=1}^n g( r_{1i}, r_{2i})l_i+
    \bar{\lambda}\sum_{i=1}^n g(r_{1i}',r_{2i}')l_i\label{eqn:g}\\
    &\leq \lambda\sum_{i=0}^{n-1}E_i+\bar{\lambda}\sum_{i=0}^{n-1}E_i\\
    &=\sum_{i=0}^{n-1}E_i
\end{align}
Therefore, the energy causality constraint is satisfied for any
$\lambda\in [0,1]$, and the new policy is energy-feasible.
Any point on the line between $(B_1,B_2)$ and
$(B_1', B_2')$ can be achieved. When $\lambda\neq 0,1$, the
inequality in (\ref{eqn:g}) is strict. Therefore, we save some
amount of energy under the new policy, which can be used to
increase the throughput for both users. This implies that
$\mathcal{D}(T)$ is strictly convex.
\end{Proof}

\begin{figure}[p]
\begin{center}
\scalebox{0.6} {\epsffile{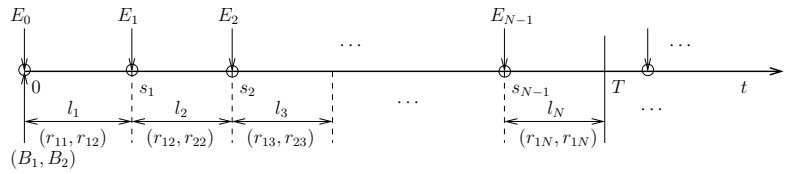}}
\end{center}
\vspace*{-0.1in}
\caption{Rates $(r_{1n},r_{2n})$ and corresponding durations $l_n$ with a given deadline $T$.}
\label{fig:bc_energy_rate}
\vspace*{-0.1in}
\end{figure}

In order to simplify the notation, in this section, for any
given $T$, we assume that there are $N-1$ energy arrival epochs
(excluding $t=0$) over $(0,T)$. We denote the last energy
arrival epoch before $T$ as $s_{N-1}$, and $s_N=T$, with
$l_N=T-s_{N-1}$, as shown in Fig.~\ref{fig:bc_energy_rate}.

Since $\mathcal{D}(T)$ is a strictly convex region, its
boundary can be characterized by solving the following
optimization problem for all $\mu_1, \mu_2 \geq 0$,
\begin{eqnarray}
\max_{\rv_1,\rv_2} & &\mu_1 \sum_{n=1}^N r_{1n}l_n+\mu_2\sum_{n=1}^N r_{2n}l_n \nonumber \\
\mbox{s.t.} & & \sum_{n=1}^j g(r_{1n},r_{2n})l_n\leq\sum_{n=0}^{j-1} E_{n}, \quad
\forall j: 0< j\leq N\label{prob1}
\end{eqnarray}
where $l_n$ is the length of the duration between two
consecutive energy arrival instances $s_{n}$ and $s_{n-1}$,
i.e., $l_n=s_n-s_{n-1}$, and $\rv_1$ and $\rv_2$ denote the
rate sequences $r_{1n}$ and $r_{2n}$ for users 1 and 2,
respectively. The problem in (\ref{prob1}) is a convex
optimization problem with a convex cost function and a convex
constraint set, therefore, the unique global solution should
satisfy the extended KKT conditions.

\begin{figure}[p]
\begin{center}
\scalebox{0.65} {\epsffile{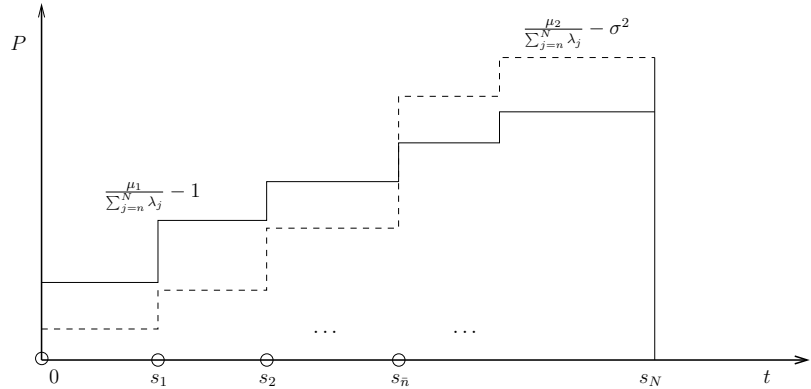}}
\end{center}
\vspace*{-0.1in}
\caption{The value of the optimal transmit power is always equal to the curve on top.}
\label{fig:power_curve}
\vspace*{-0.1in}
\end{figure}

The Lagrangian is
\begin{align}
\mathcal{L}(\rv_1,\rv_2,\lambdav,\gammav)=&\mu_1 \sum_{n=1}^N r_{1n}l_n+\mu_2\sum_{n=1}^N r_{2n}l_n
\nonumber\\
&-\sum_{j=1}^N\lambda_j\left(\sum_{n=1}^j g(r_{1n},r_{2n})l_n-\sum_{n=0}^{j-1} E_{n}\right)+
\sum_{n=1}^N\gamma_{1n} r_{1n}+\sum_{n=1}^N \gamma_{2n}r_{2n}
\end{align}
Taking the derivatives with respect to $r_{1n}$ and
$r_{2n}$, and setting them to zero, we have
\begin{align}
  \mu_1+\gamma_{1n}-\left (\sum_{j=n}^N\lambda_j\right)2^{2(r_{1n}+r_{2n})}&=0, \quad n=1,\ldots, N\label{cond1}\\
  \mu_2+\gamma_{2n}- \left(\sum_{j=n}^N\lambda_j\right)\Big(2^{2(r_{1n}+r_{2n})}+
  (\sigma^2-1)2^{2r_{2n}}\Big)&=0, \quad n=1,\ldots, N\label{cond2}
\end{align}
where $\gamma_{1n}=0$ if $r_{1n}>0$, and $\gamma_{2n}=0$ if
$r_{2n}>0$. Based on these KKT optimality conditions, we first
prove an important property of the optimal policy.
\begin{Lemma}\label{lemma4}
The optimal total transmit power of the transmitter is
independent of the values of $\mu_1,\mu_2$, and it is the same
as the single-user optimal transmit power. Specifically,
\begin{align}
i_n&= \arg \min_{\substack{i_{n-1}< i\leq N}}
\left\{\frac{\sum^{i-1}_{j=i_{n-1}}E_j}{s_i-s_{i_{n-1}}}\right\}\\
P_n&=\frac{\sum^{i_n-1}_{j=i_{n-1}}E_j}{s_{i_n}-s_{i_{n-1}}}\label{eqn:p_min}
\end{align}
i.e., at $t=s_{i_n}$, $P_n$ switches to $P_{n+1}$.
\end{Lemma}
\begin{Proof}
Based on the expression of $g(r_{1n},r_{2n})$ in (\ref{dfn:g})
and the KKT conditions in (\ref{cond1})-(\ref{cond2}), we have
\begin{align}
   g(r_{1n},r_{2n})&=\frac{\mu_2+\gamma_{2n}}{\sum_{j=n}^N\lambda_j}-\sigma^2\label{eqn:g1}\\
   &\geq 2^{2(r_{1n}+r_{2n})}-1\label{p_ineq}\\
   &=\frac{\mu_1+\gamma_{1n}}{\sum_{j=n}^N\lambda_j}-1\\
   &\geq \frac{\mu_1}{\sum_{j=n}^N\lambda_j}-1\label{eqn:g3}
\end{align}
where (\ref{p_ineq}) becomes an equality when $r_{2n}=0$.
Therefore, when $r_{2n}>0$, (\ref{eqn:g1})-(\ref{eqn:g3}) imply
\begin{align}\label{g1}
     g(r_{1n},r_{2n})&=\frac{\mu_2}{\sum_{j=n}^N\lambda_j}-\sigma^2>\frac{\mu_1}{\sum_{j=n}^N\lambda_j}-1
\end{align}

When $r_{2n}=0$, we must have $r_{1n}>0$. Otherwise, if
$r_{1n}=0$, we can always let the weaker user transmit with
some power over this duration without contradicting with any
energy constraints. Since there is no interference from the
stronger user, the departure from the weaker user can be
improved, thus it contradicts with the optimality of the
policy. Therefore, when $r_{2n}=0$, $\gamma_{1n}=0$, and
(\ref{eqn:g1})-(\ref{eqn:g3}) imply
\begin{align}\label{g2}
  g(r_{1n},r_{2n})&= \frac{\mu_1}{\sum_{j=n}^N\lambda_j}-1>\frac{\mu_2}{\sum_{j=n}^N\lambda_j}-\sigma^2
\end{align}

Therefore, we can express $g(r_{1n},r_{2n})$ in the following way:
\begin{align}\label{g3}
  g(r_{1n},r_{2n})&=\max\left\{ \frac{\mu_1}{\sum_{j=n}^N\lambda_j}-1, \frac{\mu_2}{\sum_{j=n}^N\lambda_j}-\sigma^2\right\}
\end{align}
Plotting these two curves in Fig.~\ref{fig:power_curve}, we
note that the optimal transmit power, $P_n=g(r_{1n},r_{2n})$,
is always the curve on the top. If
$\frac{\mu_2}{\sum_{j=n}^N\lambda_j}-\sigma^2>\frac{\mu_1}{\sum_{j=n}^N\lambda_j}-1$
for some $\bar{n}$, then, we have
\begin{align}
  \frac{\mu_2-\mu_1}{\sum_{j=n}^N\lambda_j}\geq\frac{\mu_2-\mu_1}{\sum_{j=\bar{n}}^N\lambda_j}>\sigma^2-1, \quad \forall n>\bar{n}
\end{align}
where the first inequality follows from the KKT condition that $\lambda_j\geq 0$ for $j=1,2,\ldots N$. Therefore, we conclude that there exists an integer
$\bar{n}$, $0\leq \bar{n}\leq N$, such that, when $n\leq
\bar{n}$,  $r_{2n}=0$; and when $n> \bar{n}$, $r_{2n}>0$.

Furthermore, (\ref{g1})-(\ref{g2}) imply that, the energy
constraint at $t=s_{\bar{n}}$ must be tight. Otherwise,
$\lambda_{\bar{n}}=0$, and (\ref{g2}) implies
\begin{align}
     g(r_{1\bar{n}},r_{2\bar{n}})&= \frac{\mu_1}{\sum_{j=\bar{n}+1}^N\lambda_j}-1>\frac{\mu_2}{\sum_{j=\bar{n}+1}^N\lambda_j}-\sigma^2= g(r_{1,\bar{n}+1},r_{2,\bar{n}+1})
\end{align}
which contradicts with (\ref{g1}). Therefore, in the following,
when we consider the energy constraints, we only need to
consider two segments $[0,s_{\bar{n}})$ and
$[s_{\bar{n}+1},s_N)$ separately.

When $n< \bar{n}$, based on (\ref{g1}), if $\lambda_n=0$, we
have $g(r_{1n},r_{2n})=g(r_{1,n+1},r_{2,n+1})$. Starting from
$n=1$, $ g(r_{1n},r_{2n})$ remains a constant until an energy
constraint becomes tight. Therefore, between any two
consecutive epochs, when the energy constraints are tight, the
power level remains constant. Similar arguments hold when
$n\geq \bar{n}$. Thus, the corresponding power level is
\begin{align}
     P_n&=\frac{\sum^{i_n-1}_{j=i_{n-1}}E_j}{s_{i_n}-s_{i_{n-1}}}
\end{align}
where $s_{i_{n-1}}$ and $s_{i_n}$ are two consecutive epochs
with tight energy constraint.

Finally, we need to determine the epochs when the energy
constraint becomes tight.  Another observation is that
$g(r_{1\bar{n}},r_{2\bar{n}})$ must monotonically increase in
$n$, as shown in Fig.~\ref{fig:power_curve}. This is because both of these two curves monotonically increase, and the maximum value of these two curves should monotonically increase also. Therefore, based on the monotonicity
of the transmit power, we conclude that
\begin{align}
    i_n&= \arg \min_{\substack{i_{n-1}< i\leq N}}\left\{\frac{\sum^{i-1}_{j=i_{n-1}}E_j}{s_i-s_{i_{n-1}}}\right\}
\end{align}
This completes the proof.
\end{Proof}

Since the power can be obtained directly irrespective of the
values of $\mu_1$, $\mu_2$, the optimization problem in
(\ref{prob1}) is separable over each duration $[s_{n-1},s_n)$.
Specifically, for $0<n\leq N$, the local optimization becomes
\begin{eqnarray}
\max_{r_{1n},r_{2n}} & &\mu_1 r_{1n}+\mu_2 r_{2n} \nonumber \\
\mbox{s.t.} & &  g(r_{1n},r_{2n})\leq P_n\label{prob2}
\end{eqnarray}
We relax the power constraint to be an inequality to make the
constraint set convex. Thus, this becomes a convex optimization
problem. This does not affect the solution since the objective
function is always maximized on the boundary of its constraint
set, i.e., the capacity region defined by the transmit power
$P_n$.

When $\frac{\mu_2}{\mu_1}\leq \frac{P_n+1}{P_n+\sigma^2}$, the
solution to (\ref{prob2}) can be expressed as
\begin{align}
  r_{1n}&=\frac{1}{2}\log_2(1+P_n)\\
  r_{2n}&=0
\end{align}
In this scenario, all of the power $P_n$ is allocated to the first user.

When $\frac{1+P_n}{\sigma^2 + P_n}\leq \frac{\mu_2}{\mu_1}\leq \sigma^2$, we have
\begin{align}
  r_{1n}&=\frac{1}{2}\log_2\left(\frac{\mu_1(\sigma^2-1)}{\mu_2-\mu_1}\right)\\
  r_{2n}&=\frac{1}{2}\log_2\left(\frac{(\mu_2-\mu_1)(P_n+\sigma^2)}{\mu_2(\sigma^2-1)}\right)
\end{align}
In this scenario, a constant amount of power, $\frac{\mu_1(\sigma^2-1)}{\mu_2-\mu_1}-1$, is allocated to the first user, and the remaining power is allocated to the second user.

When $ \frac{\mu_2}{\mu_1}> \sigma^2$, we have
\begin{align}
  r_{1n}&=0\\
  r_{2n}&=\frac{1}{2}\log_2\left(1+\frac{P_n}{\sigma^2}\right)
\end{align}
In this scenario, all of the $P_n$ is allocated to the second user.

Let us define a constant power level as
\begin{align}
P_c&=\left(\frac{\mu_1(\sigma^2-1)}{\mu_2-\mu_1}-1\right)^+
\end{align}
Based on the solution of the local optimization problem in
(\ref{prob2}), we establish another important property of the
optimal policy as follows.
\begin{Lemma}\label{lemma:cutoff}
For fixed $\mu_1$, $\mu_2$, under the optimal power policy,
there exists a constant {\it cut-off} power level, $P_c$, for
the first user. If the total power level is below this {\it
cut-off} power level, then, all the power is allocated to the
first user; if the total power level is higher than this level,
then, all the power above this {\it cut-off} level is allocated
to the second user.
\end{Lemma}

\begin{figure}[p]
\begin{center}
\scalebox{0.6} {\epsffile{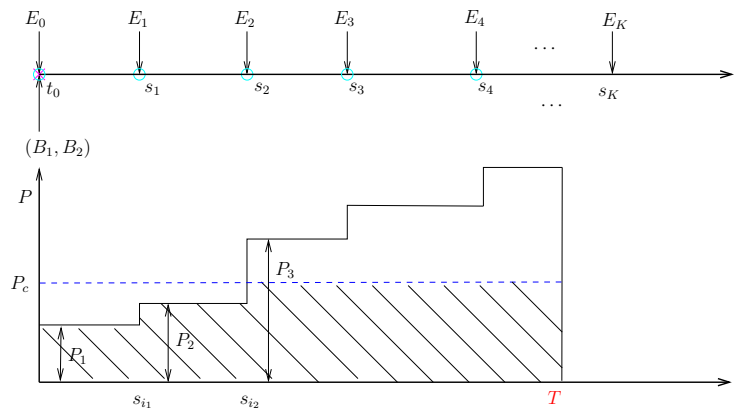}}
\end{center}
\vspace*{-0.1in}
\caption{Optimally splitting the total power between the signals that go to the two users.}
\label{fig:power}
\vspace*{-0.1in}
\end{figure}

In the proof of Lemma~\ref{lemma4}, we note that the optimal
power $P_n$ monotonically increases in $n$. Combining Lemma~\ref{lemma4}
and Lemma~\ref{lemma:cutoff}, we illustrate the
structure of the optimal policy in Fig.~\ref{fig:power}.
Moreover, the optimal way of splitting the power in each epoch
is such that both users' shares of the power monotonically
increase in time. In particular, the second user's share is
monotonically increasing in time. Hence, the path followed in
the $(B_1,B_2)$ plane is such that it changes direction to get
closer to the second user's departure axis as shown in
Fig.~\ref{fig:trajectory}. The dotted trajectory cannot be
optimal, since the path does not get closer to the second
user's departure axis in the middle (second) power epoch.

\begin{Corollary}\label{lemma2}
Under the optimal policy, the transmission rate for the first
user, $\{r_{1n}\}_{n=1}^N$, is either a constant sequence (zero
or a positive constant), or an increasing sequence. Moreover,
before $r_{1n}$ achieves its final constant value, $r_{2n}=0$;
and when $r_{1n}$ becomes a constant, $r_{2n}$ monotonically
increases in $n$.
\end{Corollary}

Based on Lemma~\ref{lemma4}, we observe that for fixed $T$,
$\mu_1$ and $\mu_2$, the optimal {\it total} power allocation
is unique, i.e., does not depend on $\mu_1$ and $\mu_2$.
However, the way the total power is split between the two users
depends on $\mu_1$, $\mu_2$. In fact, the {\it cut-off} power
level $P_c$ varies depending on the value of $\mu_2/\mu_1$.
Therefore, for different values of $\mu_2/\mu_1$, the optimal
policy achieves different boundary points on the maximum
departure region, and varying the value of $\mu_2/\mu_1$ traces
the boundary of this region.

In this section, we characterized the maximum departure region
for any given time $T$. We proved that the optimal total
transmit power is the same as in the single-user case, and
there exists a cut-off power for splitting the total transmit
power to both users. In the next section, we will use these
structural properties to solve the transmission completion
minimization problem.

\section{Minimizing the Transmission Completion Time $T$ for a Given $(B_1,B_2)$}
In this section, our goal is to minimize the transmission
completion time of both users for a given $(B_1,B_2)$. The
optimization problem can be formulated as
\begin{eqnarray}
\min_{\rv_1, \rv_2} & &T \nonumber \\
\mbox{s.t.}
& & \sum_{n=1}^j g(r_{1n},r_{2n})l_n\leq\sum_{n=1}^{j-1} E_{n}, \quad \forall j: 0<j\leq N(T)\nonumber\\
& &\sum_{n=1}^{N(T)}r_{1n}l_n= B_{1}, \quad \sum_{n=1}^{N(T)}r_{1n}l_n= B_{2}
\label{opt_prob3}
\end{eqnarray}
where $N(T)-1$ is the number of energy arrival epochs (excluding
$t=0$) over $(0,T)$, and $l_{N(T)}=T-s_{N(T)-1}$. Since $N(T)$
depends on $T$, the optimization problem in (\ref{opt_prob3})
is not a convex optimization problem in general. Therefore, we
cannot solve it using standard convex optimization tools.

We first note that this is exactly the ``dual'' problem of
maximizing the departure region for fixed $T$. They are
``dual'' in the sense that, if the minimal transmission
completion time for $(B_1,B_2)$ is $T$, then $(B_1,B_2)$ must
lie on the boundary of $\mathcal{D}(T)$, and the transmission
policy should be exactly the same for some $(\mu_1,\mu_2)$.
This is based on the fact the $\mathcal{D}(T)\subset
\mathcal{D}(T')$ for any $T< T'$. Assume $(B_1,B_2)$ does not
lie on the boundary of $\mathcal{D}(T)$. Then, either
$(B_1,B_2)$ cannot be achieved by $T$ or $(B_1,B_2)$ is
strictly inside $\mathcal{D}(T)$ and hence $(B_1,B_2)$ can be
achieved by $T'<T$. Therefore, if $(B_1,B_2)$ does not lie on
the boundary of $\mathcal{D}(T)$, then $T$ cannot be the
minimum transmission completion time.

We have the following lemma.
\begin{Lemma}\label{lemma3}
When $B_1,B_2\neq 0$, under the optimal policy, the
transmissions to both users must be finished at the same time.
\end{Lemma}
\begin{Proof}
This lemma can be proved based on Corollary~\ref{lemma2}. If
the transmission completion time for both users is not the
same, then over the last duration, we transmit only to one of
the users, while the transmission rate to the other user is
zero. This contradicts with the monotonicity of the
transmission rates for both users. Therefore, under the optimal
policy, the transmitter must finish transmitting to both users
at the same time.
\end{Proof}

This lemma is proved in \cite{uysal_paper} also, by using a
different approach. The authors prove it in \cite{uysal_paper}
mainly based on the convexity of the capacity region of the
broadcast channel.

For fixed $(B_1,B_2)$, let us denote the transmission
completion time for the first and second user, by $T_1$ and
$T_2$, respectively. We note that $T_1$ and $T_2$ depend on the
selection of the {\it cut-off} power level, $P_{c}$. In
particular, $T_1$ is monotonically decreasing in $P_c$, and
$T_2$ is monotonically increasing in $P_c$. Based on
Lemma~\ref{lemma3}, the problem of optimal selection of $P_c$,
can be viewed as solving a {\it fixed point} equation. In
particular, $P_c$ must be chosen such that, the resulting $T_1$
equals $T_2$. Therefore, we propose the following algorithm to
solve the transmission completion time, $T$, minimization
problem. Our basic idea is to try to identify the {\it cut-off}
power level $P_c$ in an efficient way.

Since the power allocation is similar to the single-user case
(c.f. Lemma \ref{lemma4}), our approach to find $T$ will be
similar to the method in \cite{ciss10, tcom-submit}. First, we
aim to identify $P_1$, the first total transmit power starting from
$t=0$ in the system. This is exactly the same as identification
of $P_1$ in the corresponding single-user problem. For this, as
in \cite{ciss10, tcom-submit}, we treat the energy arrivals as
if they have arrived at time $t=0$, and obtain a lower bound
for the transmission completion time as in \cite{ciss10,
tcom-submit}. In order to do that, first, we compute the amount of energy
required to finish $(B_1,B_2)$ by $s_1$. This is equal to
$g\left(\frac{B_1}{s_1}, \frac{B_2}{s_1}\right)s_1$, denoted as
$A_1$. Then, we compare $A_1$ with $E_0$. If $E_0$ is greater
than $A_1$, this implies that the transmitter can finish the
transmission before $s_1$ with $E_0$, and future energy
arrivals are not needed. In this case, the minimum transmission
completion time is the solution of the following equation
\begin{align}
  g\left(\frac{B_1}{T}, \frac{B_2}{T}\right)T&=E_0
\end{align}
If $A_1$ is greater than $E_0$, this implies that the final
transmission completion time is greater than $s_1$, and some of the future
energy arrivals must be utilized to complete the transmission.
We calculate the amount of energy required to finish
$(B_1,B_2)$ by $s_2$, $s_3$, \ldots, and denote them as $A_2$, $A_3$,
\ldots, and compare these with $E_0+E_1, \sum_{j=0}^2 E_j,
\sum_{j=0}^3 E_j$, \dots, until the first $A_i$ that becomes
smaller than $\sum_{j=0}^{i-1} E_j$. We denote the
corresponding time index as $\tilde{i}_1$. Then, we assume that
we can use $\sum_{i=0}^{\tilde{i}_1-1} E_i$ to transmit
$(B_1,B_2)$ at a constant rate. And, the corresponding
transmission completion time is the solution of the following
equation
\begin{align}
  g\left(\frac{B_1}{T}, \frac{B_2}{T}\right)T&=\sum_{i=0}^{\tilde{i}_1-1} E_i
\end{align}

\begin{figure}[p]
\begin{center}
\scalebox{0.65} {\epsffile{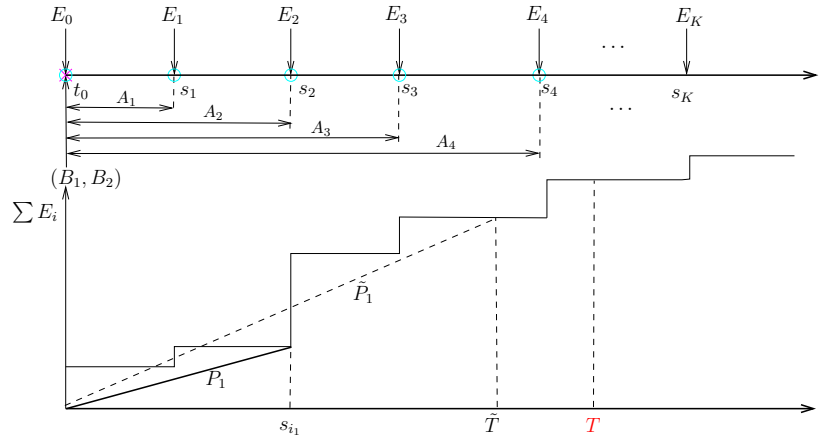}}
\end{center}
\vspace*{-0.1in}
\caption{Determining the optimal total power level of the first epoch.}
\label{fig:bc_algr}
\vspace*{-0.1in}
\end{figure}

We denote the solution to this equation as $\tilde{T}$, and the
corresponding power as $\tilde{P}_1$. From our analysis, we
know that the solution to this equation is the minimum possible
transmission completion time we can achieve. Then, we check
whether this constant power $\tilde{P}_1$ is feasible, when the
actual energy arrival times are imposed. If it is feasible, it
gives us the minimal transmission completion time; otherwise,
we get $P_1$ by selecting the minimal slope according to
(\ref{eqn:p_min}). That is to say, we draw all of the lines
from $t=0$ to the corner points of the energy arrival instances
before $\widetilde{T}$, and choose the line with the smallest
slope. We denote by $s_{i_1}$ the corresponding duration
associated with $P_1$. This is shown in Fig.~\ref{fig:bc_algr}.

Once $P_1$ is selected, we know that it is the optimal total
transmit power in our broadcast channel problem. Next, we need
to divide this total power between the signals transmitted to
the two users. Based on Lemma~\ref{lemma:cutoff} and
Corollary~\ref{lemma2}, if the {\it cut-off} power level $P_c$
is higher than $P_1$, then, the transmitter spends all $P_1$
for the stronger user; otherwise, the first user finishes its
transmission with a constant power $P_c$.

We will first determine whether $P_c$ lies in $[0,P_1]$ or it
is higher than $P_1$. Assume $P_c=P_1$. Therefore, the
transmission completion time for the first (stronger) user is
\begin{align}
T_1&=\frac{B_1}{f(P_1)}
\end{align}
Once $P_c$ is fixed, we can obtain the minimum transmission
completion time for the second user, $T_2$, by subtracting the
energy consumed by the first user, and treating $P_1$ as an
interference for the second user. This reduces the problem to
the single-user problem for the second user with fading, where
the fading level is $P_1+\sigma^2$ over $[0,T_1)$, and
$\sigma^2$ afterwards. The single-user problem with fading is
studied in \cite{energy_fading}. Since obtaining the minimal
transmission completion time is not as straightforward for the
fading channel, a more approachable way is to calculate the
maximum number of bits departed from the second user by $T_1$,
denoted as $D_2(T_1,P_c)$. In order to do that, we first
identify the optimal power allocation policy with fixed
deadline $T_1$. This can be done according to
Lemma~\ref{lemma4}. Assume that the optimal power allocation
gives us $P_1,P_2,\ldots, P_{N(T_1)}$. Then, we allocate $P_1$
to the first user over the whole duration, and allocate the
remaining power to the second user. Based on (\ref{eqn:rate2}),
we calculate the transmission rate for the second user over
each duration, and obtain $D_2(T_1,P_c)$ according to
\begin{align}
  D_2(T_1,P_c) = \sum_{i=1}^{N(T_1)} \frac{1}{2}\log\left(1 + \frac{P_n-P_c}{P_c + \sigma^2}\right)
  (s_{i_n}-s_{i_{n-1}})
\end{align}
We observe that, given $P_c$, $D_2(T_1,P_c)$ is a monotonically
increasing function of $T_1$. Moreover, given $T_1$,
$D_2(T_1,P_c)$ is a monotonically decreasing function of $P_c$.

If $D_2(T_1,P_c)$ is smaller than $B_2$, it implies that
$T_1<T_2$, and we need to decrease the rate for the first user
to make $T_1$ and $T_2$ equal. Based on
Lemma~\ref{lemma:cutoff}, this also implies that the
transmission power for the first user is a constant $P_c<P_1$.
In particular, $P_c$ is the unique solution of
\begin{align} \label{eq}
 B_2 = D_2\left(\frac{B_1}{f(P_c)}, P_c\right)
\end{align}
Note that $D_2\left(\frac{B_1}{f(P_c)}, P_c\right)$ is a
continuous, strictly monotonically decreasing function
of $P_c$, hence the solution for $P_c$ in (\ref{eq}) is unique.
Since $T_1$ is a decreasing function of $P_c$ and
$D_2\left(\frac{B_1}{f(P_c)}, P_c\right)$ is a decreasing
function of $P_c$, we can use the bisection method to solve
(\ref{eq}).  In
this case, the minimum transmission completion time is $T =
\frac{B_1}{f(P_c)}$.

\begin{figure}[p]
\begin{center}
\scalebox{0.65} {\epsffile{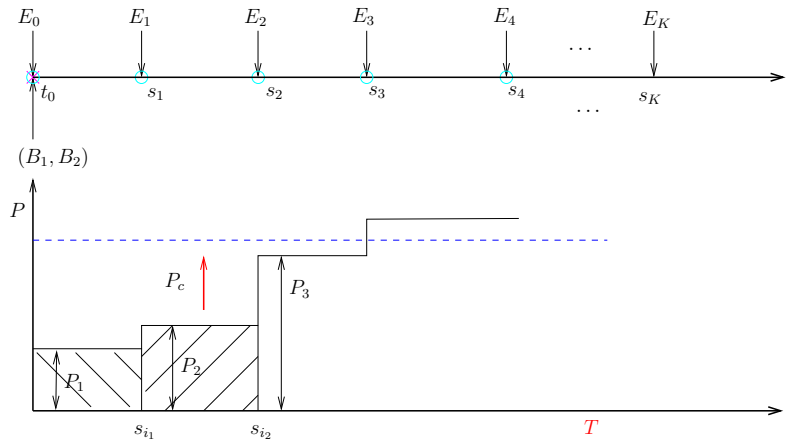}}
\end{center}
\vspace*{-0.1in}
\caption{Search for the cut-off power level $P_c$ iteratively.}
\label{fig:search_pc}
\vspace*{-0.1in}
\end{figure}

If $D_2(T_1,P_c)$ is larger than $B_2$, that implies $T_2<T_1$,
and we need to increase the power allocated for the first user
to make $T_1$ and $T_2$ equal, i.e., $P_c>P_1$. Therefore, from
Lemma~\ref{lemma:cutoff}, over the duration $[0,s_{i_1})$, the
optimal policy is to allocate the entire $P_1$ to the first
user only. We allocate $P_1$ to the first user, calculate the
number of bits departed for the first user, and remove them
from $B_1$. This simply reduces the problem to that of
transmitting $(B_1', B_2)$ bits starting at time $t=s_{i_1}$,
where $B_1'=B_1-f(P_1)s_{i_1}$. The process is illustrated in
Fig.~\ref{fig:search_pc}. Then, the minimum transmission
completion time is
\begin{align}
T =  s_{i_K} + \frac{B_1-\sum_{i=1}^Kf(P_k)(s_{i_k}-s_{i_{k-1}})}{f(P_c)}
\end{align}
where $K$ is the number of recursions needed to get $P_c$.

In both scenarios, we reduce the problem into a simple form,
and obtain the final optimal policy. Before we proceed to prove
the optimality of the algorithm, we introduce the following
lemma first, which is useful in the proof of the optimality of
the algorithm.
\begin{Lemma}\label{lemma:mono}
$f(E/T)T$ monotonically increases in $T$; $f\left(\frac{\alpha
E/T}{(1-\alpha E/T)+\sigma^2}\right)T$ monotonically increases
in $T$ also.
\end{Lemma}
\begin{Proof}
The monotonicity of both functions can be verified by taking
derivatives,
\begin{align}
   (f(E/T)T)'&=f(E/T)-\frac{E}{(2\ln2) (T+E)}
\end{align}
and
\begin{align}
   (f(E/T)T)''&=\frac{E}{2\ln2}\left(\frac{1}{(T+E)^2}-\frac{1}{T(T+E)}\right)<0
\end{align}
where the inequality follows since $E>0$. Therefore, $f(E/T)T$
is a strictly concave function, and its first derivative
monotonically decreases when $T$ increases. Since when
$\lim_{T\rightarrow \infty} (f(E/T)T)'\break=0$, when
$T<\infty$, we have $(f(E/T)T)'>0$, therefore, the monotonicity
follows.

Similarly, we have
  \begin{align}
   \left( f\left(\frac{\alpha E/T}{(1-\alpha E/t)+\sigma^2}\right)T\right)'=&\frac{1}{2}\log_2
   \left(\sigma^2+E/T\right)-\frac{1}{2}\log_2\left(\sigma^2+(1-\alpha)
   E/T\right)\nonumber\\
   &-\frac{E}{2\ln2}\frac{E}{E+\sigma^2T}+\frac{E}{2\ln2}\frac{(1-\alpha)E}{(1-\alpha)E+\sigma^2T}
   \end{align}
and
   \begin{align}
    \left( f\left(\frac{\alpha E/T}{(1-\alpha E/t)+\sigma^2}\right)T\right)''=&\frac{E^2}{2T\ln2}\left(\frac{1}{(\sigma^2T/(1-\alpha)+E)^2}-\frac{1}{(\sigma^2T+E)^2}\right)<0
  \end{align}
Again, the concavity implies that the first derivative is
positive when $T<\infty$, and the monotonicity follows.
\end{Proof}

\begin{Theorem}
The algorithm is feasible and optimal.
\end{Theorem}
\begin{Proof}
We first prove the optimality. In order to prove that the
algorithm is optimal, we need to prove that $P_1$ is optimal.
Once we prove the optimality of $P_1$, the optimality of $P_2$,
$P_3$, $\ldots$ follows. Since the solution obtained using our
algorithm always has the optimal structure described in
Lemma~\ref{lemma:cutoff}, the optimality of the power
allocation also implies the optimality of the rate selection,
thus, the optimality of the algorithm follows. Therefore, in
the following, we prove that $P_1$ is optimal.

First, we note that $P_1$ is the minimal slope up to
$\widetilde{T}$. We need to prove that $P_1$ is also the minimal
slope up to the final transmission completion time, $T$. Let us
define $T'$ as follows
 \begin{align}
   T'&=\frac{\sum_{n=0}^{\tilde{i}_1} E_n}{P_1}
 \end{align}
Assume that with $\tilde{P}_1$, we allocate $\alpha
\tilde{P}_1$ to the first user, and finish $(B_1,B_2)$ using
constant rates. Then, we allocate $\alpha P_1$ to the first
user, and the rest to the second user. Based on
Lemma~\ref{lemma:mono}, we have
\begin{align}
   f(\alpha P_1)T'&\geq f(\alpha \tilde{P}_1)\tilde{T}=B_1\label{eqn:b1}\\
   f\left(\frac{\alpha P_1}{(1-\alpha )P_1+\sigma^2}\right)T'&\geq  f\left(\frac{\alpha \tilde{P}_1}{(1-\alpha )\tilde{P}_1+\sigma^2}\right)\hat{T}=B_2\label{eqn:b2}
\end{align}
Therefore, $T'$ is an upper bound for the optimal transmission
completion time. Since $P_1$ is the minimal slope up to $T'$,
we conclude that $P_1$ is optimal throughout the transmission. Following
similar arguments, we can prove the optimality of the rest of
the power allocations. This completes the proof of optimality.

In order to prove that the allocation is feasible, we need to
show that the power allocation for the first user is always
feasible in each step. Therefore, in the following, we first
prove that $P_1$ is feasible when we assume that $P_c=P_1$. The
feasibility of $P_1$ also implies the feasibility of the rest
of the power allocation. With the assumption that $P_c=P_1$,
the final transmission time for the first user is
\begin{align}
    T_1&=\frac{B_1}{f(P_1)}\leq \frac{B_1}{f(\alpha P_1)}
\end{align}
Based on (\ref{eqn:b1}) and (\ref{eqn:b2}), we know that
$T_1<T'$. Since $P_1$ is feasible up to $T'$, therefore, $P_1$ is feasible when we assume that
$P_c=P_1$. The feasibility of the rest of the power allocations
follows in a similar way. This completes the feasibility part of the proof.
\end{Proof}

\section{Numerical Examples}
\begin{figure}[p]
\begin{center}
\scalebox{0.65} {\epsffile{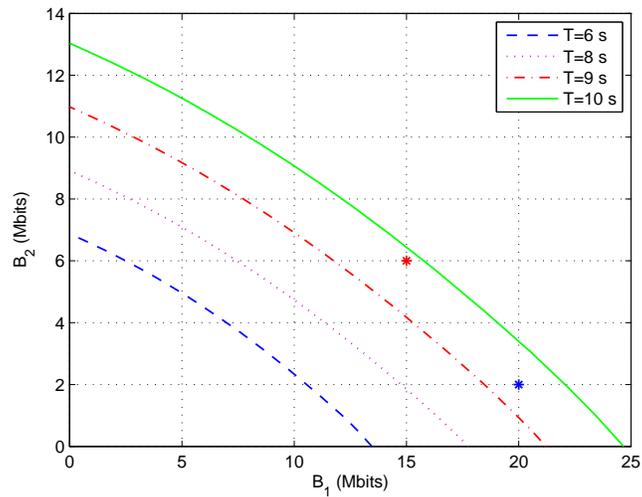}}
\end{center}
\vspace*{-0.1in}
\caption{The maximum departure region of the broadcast channel for various $T$.}
\label{fig:bc_region}
\vspace*{-0.1in}
\end{figure}

We consider a band-limited AWGN broadcast channel, with
bandwidth $W=1$ MHz and the noise power spectral density
$N_0=10^{-19}$ W/Hz. We assume that the path loss between the
transmitter and the first receiver is about $100$ dB, and the
path loss between the transmitter and the second user is about
$105$ dB. Then, we have
\begin{align}
  r_1&=W\log_2\left(1+\frac{\alpha Ph_1}{N_0W}\right)=\log_2\left(1+\frac{\alpha P}{10^{-3}}\right)\mbox{Mbps}\\
  r_2&=W\log_2\left(1+\frac{(1-\alpha) Ph_2}{\alpha Ph_2+N_0W}\right)=\log_2\left(1+\frac{(1-\alpha) P}{\alpha P+10^{-2.5}}\right)\mbox{Mbps}
\end{align}
Therefore,
\begin{align}
  g(r_1,r_2)&=10^{-3}2^{r_1+r_2}+(10^{-2.5}-10^{-3})2^{r_2}-10^{-2.5}\quad \mbox{W}
\end{align}
For the energy harvesting process, we assume that at times
$\tv=[0,2,5,6,8,9,11]$ s, we have energy harvested with amounts
$\Ev=[10,5,10,5,10,10,10]$ mJ. We find the maximum departure
region $\mathcal{D}(T)$ for $T=6,8,9,10$ s, and plot them in
Fig.~\ref{fig:bc_region}. We observe that the maximum departure region is convex for each value
of $T$, and as $T$ increases, the maximum departure region
monotonically expands.

\begin{figure}[p]
\begin{center}
\scalebox{0.65} {\epsffile{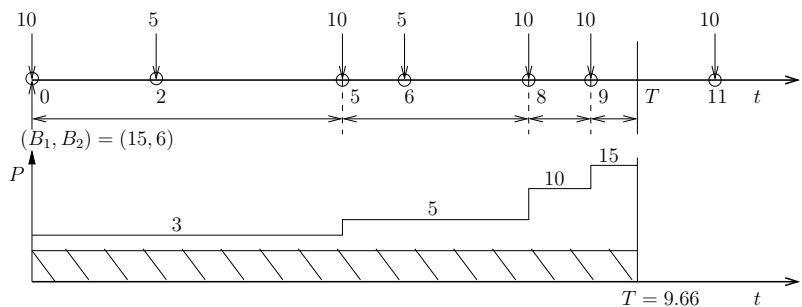}}
\end{center}
\vspace*{-0.1in}
\caption{Cut-off power $P_c=1.933$ mW. Optimal transmit rates are $r_1=1.552$ Mbps,
$\rv_2=[ 0.274,0.680,1.369, 1.834]$ Mbps,
with durations $\lv=[5,3,1,0.66]$ s.}
\label{fig:bc_example1}
\vspace*{-0.1in}
\end{figure}

Then, we aim to minimize the transmission completion time with
$(B_1,B_2)=(15,6)$ Mbits. Following our algorithm, we obtain the optimal
transmission policy, which is shown in
Fig.~\ref{fig:bc_example1}. We note that the powers change only
potentially at instances when energy arrives
(Lemma~\ref{lemma:const}); power sequence is monotonically
increasing and ``majorized'' over the whole transmission
duration (Lemma~\ref{lemma4}). We also note that, for this
case, the first user transmits at a constant rate, and the
rate for the second user monotonically increases. The transmitter finishes its transmissions to both users by time $T=9.66$ s, and the last
energy harvest at time $t=11$ s is not used.

\begin{figure}[p]
\begin{center}
\scalebox{0.65} {\epsffile{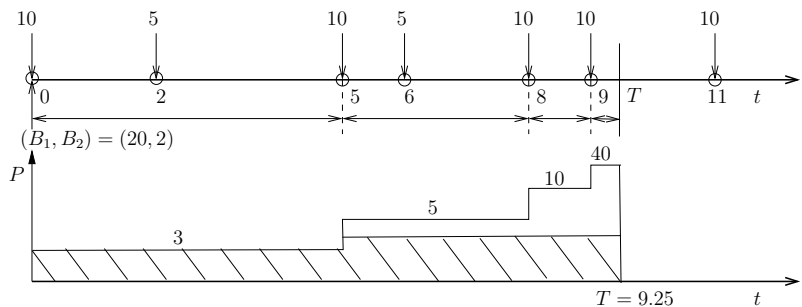}}
\end{center}
\vspace*{-0.1in}
\caption{Cut-off power $P_c=4.107$ mW. Optimal transmit rates
$\rv_1=[2, 2.353, 2.353, 2.353 ]$ Mbps and $\rv_2=[0,0.167, 0.856,2.570]$ Mbps,
with durations $\lv=[5,3,1,0.25]$ s.}
\label{fig:bc_example2}
\vspace*{-0.1in}
\end{figure}

Next, we consider the example when $(B_1,B_2)=(20,2)$ Mbits, we
have the optimal transmission policy, as shown in
Fig.~\ref{fig:bc_example2}. In this example, the cut-off power
is greater than $P_1$, and therefore, $P_1$ is allocated to the
first user only over $[0,5)$ s, and after $t=5$ s, the first user
keeps transmitting at a constant rate until all bits are
transmitted. In this case, the transmission rates for both users monotonically increase. The transmitter finishes its transmissions by time $T=9.25$ s, and the last energy harvest is not used.

\section{Conclusions}
We investigated the transmission completion time minimization
problem in an energy harvesting broadcast channel. We first
analyzed the structural properties of the optimal transmission
policy, and proved that the optimal total transmit power has
the same structure as in the single-user channel. We also
proved that there exists a {\it cut-off} power for the stronger
user. If the optimal total transmit power is lower than this
cut-off level, all power is allocated to the stronger user, and
when the optimal total transmit power is greater than this
cut-off level, all power above this level is allocated to the
weaker user. Based on these structural properties of the
optimal policy, we developed an iterative algorithm to obtain
the globally optimal off-line transmission policy.

\newpage

\end{document}